# Unsupervised Classification of Single-Molecule Data with Autoencoders and Transfer Learning

Anton Vladyka[1], Tim Albrecht[1]

[1] School of Chemistry, University of Birmingham, Edgbaston Campus, Birmingham B15 2TT, United Kingdom

E-mail: t.albrecht@bham.ac.uk

**Abstract**

Datasets from single-molecule experiments often reflect a large variety of molecular behaviour. The exploration of such datasets can be challenging, especially if knowledge about the data is limited and a priori assumptions about expected data characteristics are to be avoided. Indeed, searching for pre-defined signal characteristics is sometimes useful, but it can also lead to information loss and the introduction of expectation bias. Here, we demonstrate how Transfer Learning-enhanced dimensionality reduction can be employed to identify and quantify hidden features in single-molecule charge transport data, in an unsupervised manner. Taking advantage of open-access neural networks trained on millions of seemingly unrelated image data, our results also show how Deep Learning methodologies can readily be employed, even if the amount of problem-specific, 'own' data is limited.

Studies of molecular behaviour at the single-molecule level have become routine in many areas of physics, chemistry and biology.[1-3] Rather than focusing ensemble averages only, single-molecule data are can be 'information rich' and reflect a variety of molecular behaviour, in terms of their structural, electronic and other physical properties. For example, in charge transport studies individual molecules are usually 'wired' between two electrodes, in order to study their electric conductance or current-voltage characteristics.[4,5] However, they can bind in different geometries, at different surface sites and display complex dynamic behaviour, which is then reflected in the measured response. The presence of several molecules in the junction at the same time can give rise to additional complexity, for example multiple conductance values corresponding to one, two or more molecules wired in parallel or intermolecular interaction effects.[6-8] This complexity usually necessitates the recording of large datasets, often in an automated fashion, in order to fully capture the behaviour of the system. Hence, datasets containing tens of thousands of individual current-voltage, current-distance or current-time traces are now not uncommon. Analysing such datasets can be rather challenging, especially when knowledge of the system under study is limited or when it is undesirable for prior expectation to enter the analysis from the start. A hypothesis-driven approach, while sometimes useful, risks missing important information (e.g., the data that do not follow the initial expectation), is prone to user or confirmation bias and makes it more difficult to identify and quantify potential sub-populations in the data.[9-13] Hence, unsupervised approaches that do not require 'labelled' data and avoid these pitfalls are preferable in those instances.

Accordingly, dimensionality reduction techniques have proven to be a powerful tool in this regard. The fundamental idea here is to represent high-dimensional data by a reduced number of descriptors or classifiers that capture the data's salient features. In this reduced dimensional vector representation, similar data traces are thought to be in close proximity, whereas dissimilar traces are further apart. If several distinct point clusters emerge, they can potentially be separated and quantified using suitable clustering methods, such as k-means, Gaussian Mixture Models, density- or density-gradient-based methods, and the constituent data traces be further analysed and interpreted.[14] Thus, rather than defining an expected outcome prior to the analysis, common features in the data emerge as such during or after the clustering step.

Various dimensionality reduction techniques are available, including Principal Component Analysis (PCA) and its variations, t-distributed stochastic neighbour embedding (t-SNE), and specifically developed in the context of charge transport data, Multi-Parametric Vector Classification (MPVC), and many others.[9,15-17] Autoencoders (AE) are another method that can be employed to this effect, with close relations to PCA.[18,19] AEs are neural networks composed of multiple layers of neurons, which are first trained to represent the data in layers with a progressively smaller number of neurons/dimensions (encoder stage), to produce a low-dimensional 'code' layer, and then to reconstruct the original data, layer-by-layer with increasing number of neurons, towards the output (decoder stage). For example, in the case of tunnelling current-distance data containing 2000 data points per trace, the first input layer of the AE may contain 2000 neurons, a number that is reduced in a step-wise fashion with each layer to potentially only two or three at the 'code' stage. Extraction of the numerical output at this point thus provides a low-dimensional representation that should capture the salient features of the input data (but not every detail), and can be used for classification. However, the training of the network, and multi-layer ('deep') AEs in particular, can be difficult and optimisation via back-propagation often ends up in local minima and hence poor performance.[18,20] Accordingly, we have found in our results reported here, that an AE applied directly to the raw input data did not yield satisfactory classification results (in contrast to a recent report [21]).

The performance was enhanced significantly, however, when the AE was combined with neural networks trained on seemingly unrelated image data. As an example of Transfer Learning (TL), the network identifies characteristic features in the measured traces, even though it had not been trained on this type of data. One such neural networks we employed here was AlexNet.[22] AlexNet has been trained on ImageNet, a dataset containing millions of images manually classified into 1000 real-world classes (such as animals, constructions and vehicles). It can formally be separated into two parts, namely feature extractor and classifier. Here we only exploit the capabilities of the former, namely to recognise certain features in our charge transport data, and use the feature extractor output as input for dimensionality reduction, e.g. via an AE, figure 1. If passed on to the classification part, the network would formally identify one of the real-world classes it had been trained on, such as a nematode - but these are of no relevance in the present context.

The fact that feature recognition did not rely on training data from the actual classification task highlights an important factor, namely that training data may be "borrowed" from other contexts, where they are more abundant or easier to obtain. This significantly relaxes the requirement of obtaining large amounts of task-specific data, with a view on network training.

Hence, we have pursued several objectives in the present work, based on both simulated and experimental charge transport data.[23, 24] Firstly, we apply several dimensionality reduction techniques to the data and compare the results, noting the complexities of a like-for-like performance comparison. Secondly, we show how TL can enhance the classification performance and assess the generality of the approach. In our case, we have used three different neural networks for feature extraction, namely AlexNet,[22] the winner of the ICLR-2012 image recognition context, ResNet and VGGnet16.[25,26] Thirdly, we demonstrate for a molecular system, MBdNC, how such advanced classification methodologies can help extract additional, physically meaningful information from the data that were previously undetected, and also identify which global features in the data are most likely responsible for the classification result.

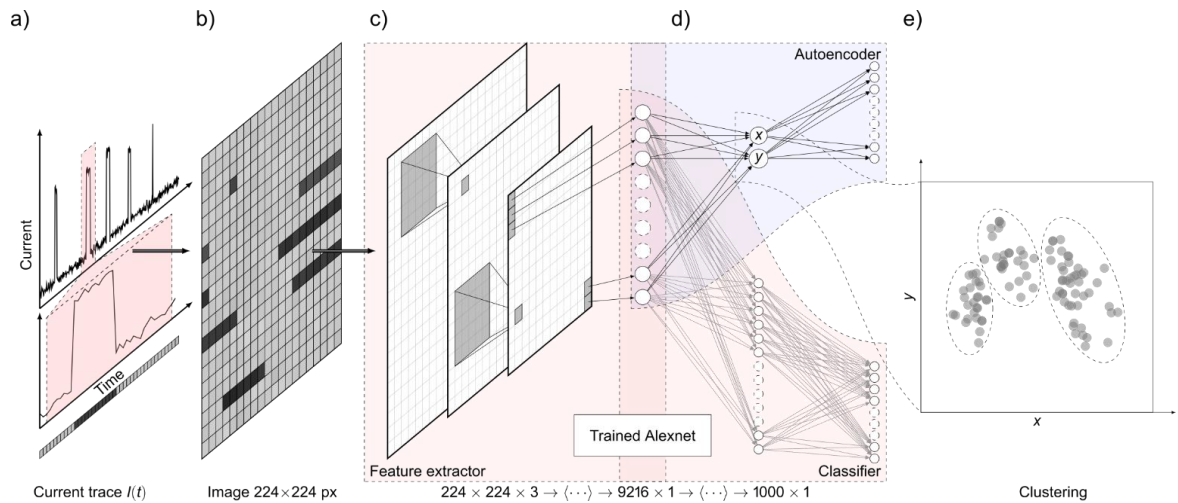

**Figure 1**. Schematics of the TL approach used here. a) The input signal and its encoding into a one-dimensional array. b) Corresponding b/w image. c) Trained AlexNet. d) Autoencoder (AE) built on the feature extractor of AlexNet. e) Two-dimensional representation of the data after dimensionality reduction using AE.

**Results and Discussion**

We first illustrate the working principle with simulated data, similar to those obtained in quantum tunnelling-based nucleotide recognition experiments.[27-32, 9, 23] In these experiments, a tunnelling junction composed of two electrodes with gap sizes in the low nanometre range and with a well-defined bias voltage are created, for example in an STM configuration in solution. If DNA nucleotides or other analytes are present and the electrode gap size is suitable, the analyte can diffuse into and temporarily bridge the electrode gap, thereby causing a measurable change in the tunnelling conductance. This leads to current-time ($I(t)$) traces that feature short-lived spikes, corresponding to individual binding events, as shown in fig. 2 a). Since different nucleotides in the electrode junction produce different event characteristics, the corresponding tunnelling modulation can be used to differentiate between the different nucleotides. Here, we employ simulated data (2000 traces per nucleotide, see Methods section for simulation parameters), to facilitate the systematic variation of the input data and the validation of the classification results.

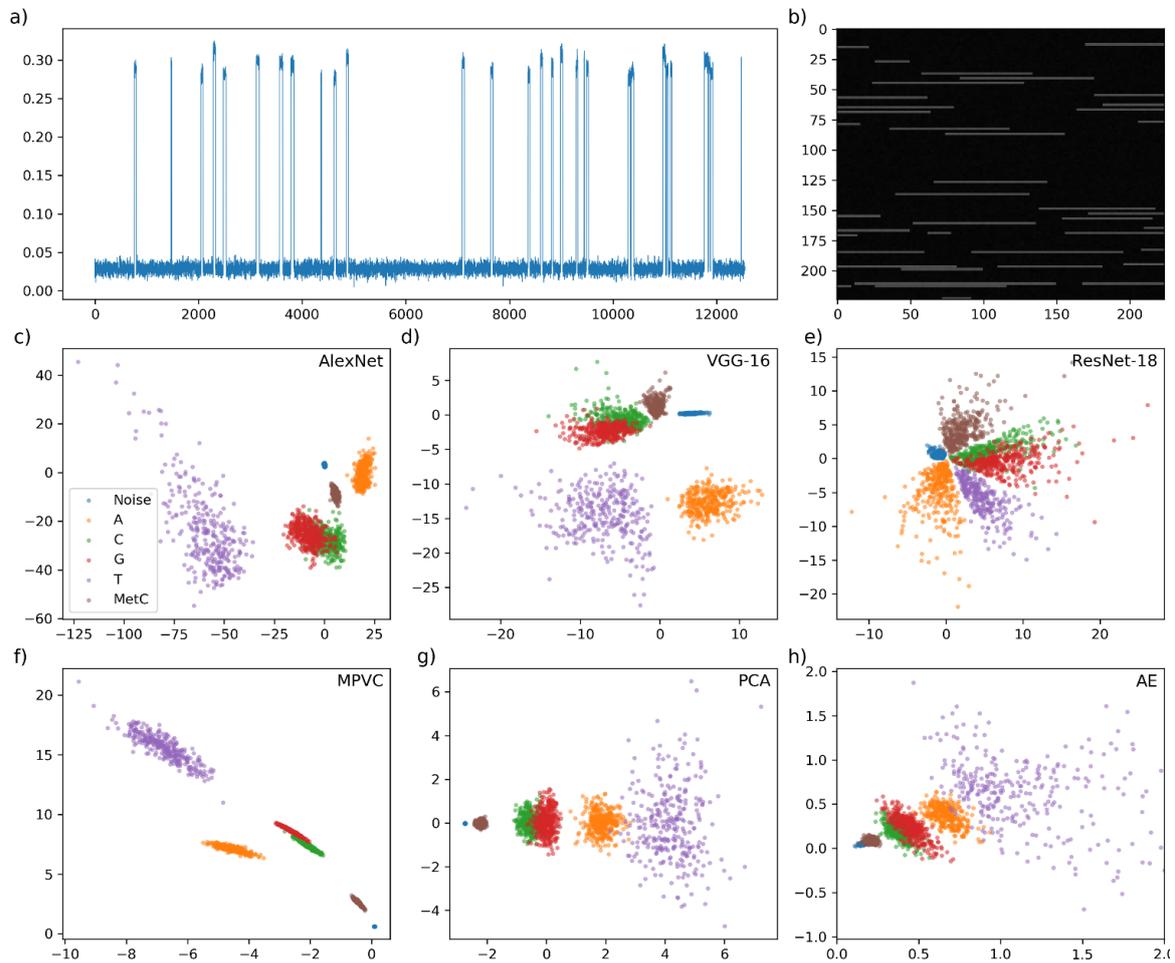

**Figure 2.** TL approach on simulated nucleotide data. a) Typical sample trace with 112x112 =12544 points. b) Corresponding 224x224 b/w image used as an input of the feature extractor for the trained neural network. c-e) Scatter plots for all 2000 simulated traces after applying AE-based dimensionality reduction approach on the features extracted using trained convolutional neural networks: AlexNet, VGG-16 and ResNet-18, respectively. f-h) Scatter plots for all 2000 simulated traces after applying 'classical; dimensionality reduction approaches: MPVC, PCA and AE, respectively. The color-coding is the same for all figures c-h).

As described in more detail in the Methods section, we converted each of the 2000 simulated I(t) traces into the corresponding images, figure 2b). These images were then passed through the (unmodified) feature extractor component of pre-trained image recognition networks AlexNet, VGG-16 and ResNet-18. In the case of AlexNet, feature extraction yielded 2000 9216-dimensional feature vectors, which were then used as input for an AE as an unsupervised dimensionality reduction method, ultimately producing a two-dimensional representation in the code layer, cf. Fig. S1 in the Supporting

Information (SI). These value pairs were then plotted in the (x,y) plane for clustering and classification, panels c-e). The same procedure was repeated several times with newly generated data as well as with the same data after re-initializing the AE. In all cases, the obtained two-dimensional cluster representation turned out to be robust in that all nucleotides, except for C and G, formed well-separated clusters. The inability to separate C and G was rooted in the fact that the simulation parameters were very similar, hence resulting in significant overlap, in terms of the event characteristics. We note that, in comparison to VGG-16 and ResNet-18, AlexNet is the least complex network and requires less computing power for feature extraction. For comparison, we also applied more classical dimensionality reduction techniques, such as an AE with the raw data as input (i.e., without feature extraction), PCA and MPVC, panels f-h). In terms of cluster separation, PCA and the 'direct' AE show similar results, even though they are performing worse than the AlexNet-based approach. At least for this dataset, MPVC appears to show the best separation, an observation that is however not borne with other datasets, as shown below.

From these initial tests, it is however apparent that the incorporation of the feature extractor lead to improved separation, compared to the 'direct' AE approach and conventional PCA, as employed here. Out of the three feature extractor networks, AlexNet displayed the best performance with least computational cost. We will therefore focus on the AlexNet/AE combination in the following, when analysing experimental molecular break-junction junction data.

To this end, we apply the TL approach to conductance-distance traces, G(d), measured using a Mechanically Controlled Break-Junction (MCBJ) in solution for three different molecules, 1,4-benzenediisocyanide (BdNC), 2,5-dimethyl-1,4-benzenediicyanide (MBdNC) and 2,5-di-tert-butyl-1,4-benzenediisocyanide (tBuBdNC).[24] In this case, each conductance trace contained 224 conductance values (in units of log($G/G_0$), where $G_0=2e^2/h$=77.5 μS is the conductance quantum). Every point of the trace corresponds to the relative displacement of the electrodes from -2 A to 20.3 Å, with a step of 0.1 Å. Overall, 1425 conductance traces were analysed, namely a mixture of 481 traces for BdNC, 379 traces for MBdNC and 565 traces for tBuBdNC.

First, we converted each trace into a 224x224 RGB image, as standard input for AlexNet. To do this, each conductance trace was reshaped into a two-dimensional array with 112x112 bins, which was then upscaled to a resolution of 224x224, figure 3a). In the next step, the images were passed through the feature extractor of AlexNet, resulting in the 9216-dimensional feature vector as above, which again served as input to the AE. The AE was trained with a learning rate of 0.0001 using an Adam optimizer.[33] Finally after training, the neural activations in the middle hidden (code) layer were extracted and used as reduced dimensional representation of the data, as shown in fig. 3. For comparison, we also employed MPVC, PCA, and direct AE, as well as t-distributed stochastic neighbour embedding (t-SNE) on the raw data, see figure S2 in the SI.

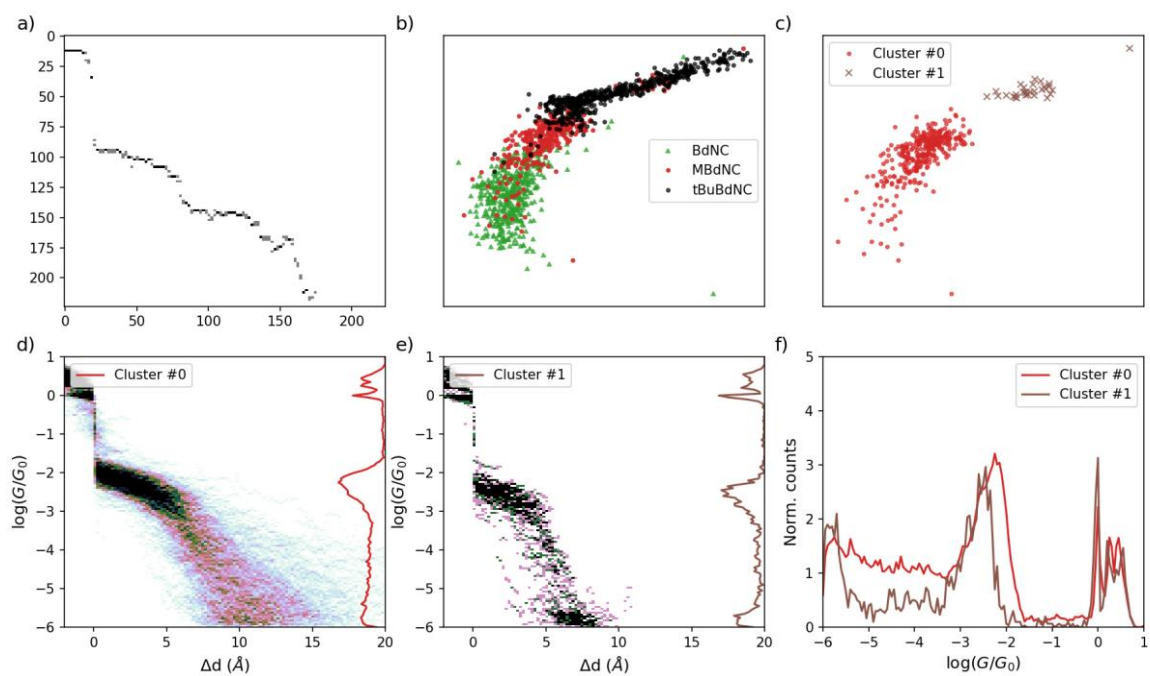

**Figure 3**. Transfer learning approach applied to molecular junction data. a) Typical conductance trace converted into 224x224 b/w image. b) AlexNet-based dimensionality reduction approach applied to the molecular junction data for three molecules. c) Two clusters for MBdNC molecule, determined by the Spectral Clustering algorithm. d-e) Combined 2D/1D conductance histograms for clusters 0 and 1, respectively. f) Conductance histograms for clusters 0 and 1.

Some observations are appropriate in relation to these results, before proceeding further with the discussion. Firstly, a detailed physical interpretation of the experimental results can be found in the original work reported in reference;[24] we will mainly focus on the data analysis aspects here. It is nevertheless important to note that conductance values below -5.5 on the logarithmic scale are at the sensitivity limit of the amplifier used. Thus, while data points in this region are "seen" by the classification algorithm, further physical interpretation is problematic. Secondly, it is worth noting that the shape of individual G(d) traces cannot be inferred from the characteristics in the 2D histogram, as the latter represents a convolution of conductance and distance information and is potentially made up of sub-populations with different event characteristics. Therefore, this aspect requires further analysis, as shown below.

In terms of the classification results, however, neither of these five approaches is able to separate all three different molecules, figures 3b) (AlexNet/AE) and S2 (MPVC, PCA, direct AE, t-SNE). While BdNC and tBuBdNC generally formed well-separated clusters, MBdNC data tended to overlap with at least one or both of the previous datasets. This broadly reflects the characteristics of the G(d) traces, including the variance inherent to all three datasets.

However, closer inspection of the individual molecular datasets lead to a rather unexpected result. While the BdNC and tBuBdNC datasets only yielded single clusters each (see however below), two separate sub-clusters emerged for MBdNC after 300 epochs of training, figure 3c). For these two subpopulations, (two-dimensional) conductance-displacement histograms and (one-dimensional) conductance histograms were calculated, figure 3d-f). In relation to the former, it is apparent that traces in cluster 0 produce a single, well-defined plateau at approximately $\log(G/G_0) \approx -2$ with significant variance during drop-off at larger displacements. The conductance-displacement histogram from cluster 1 features two high point density regions, one at $-2.5 < \log(G/G_0) < -2$ and one below $-5.5$ at larger displacements, i.e. close to sensitivity limit of the amplifier. Overall, the traces in this cluster displayed a better defined (lower variance) and shorter drop-off towards larger displacements. The high-conductance plateau in cluster 1 appears to be slightly shifted towards a lower conductance value, as shown in the one-dimensional histogram in fig. 3f).

We will return to this point later, but first discuss the effect of training on the classification. For this purpose, the training of the AE component was re-run for 10000 epochs for all three molecules and the state of the analysis tracked during the process, as in shown in Figure 4a-i for 100, 300 and 1000 epochs (no further evolution was observed for epochs > 1000).

For BdNC and MBdNC, the outcome after this renewed and extended training was the same as before, namely a single population for the former and two sub-populations for the latter molecule. For tBuBdNC, however, the dataset did progressively separate into two groups, as the number of epochs increased to 1000, and analysed as before, figure 4j-l). This time, the conductance-displacement histogram for cluster 0 showed two high point density regions, one at $\log(G/G_0) \approx -3$ and another one below $-5$ at longer displacements, panel j). On the contrary, the conductance-displacement histogram determined from cluster 1 only contained one apparent high point density region at approximately $-3$, but none at longer displacements, panel k). Notably, the (1D) conductance histograms for clusters 0 and 1 showed the main conductance peaks at $\log(G/G_0) \approx -3.4$, i.e. with similar shape and location (even though the peak from cluster 1 appears to be somewhat narrower).

As for MBdNC above, the main apparent differences between the 2D histograms again appear to be the different steepness of the decay and the high point density region at low conductance values.

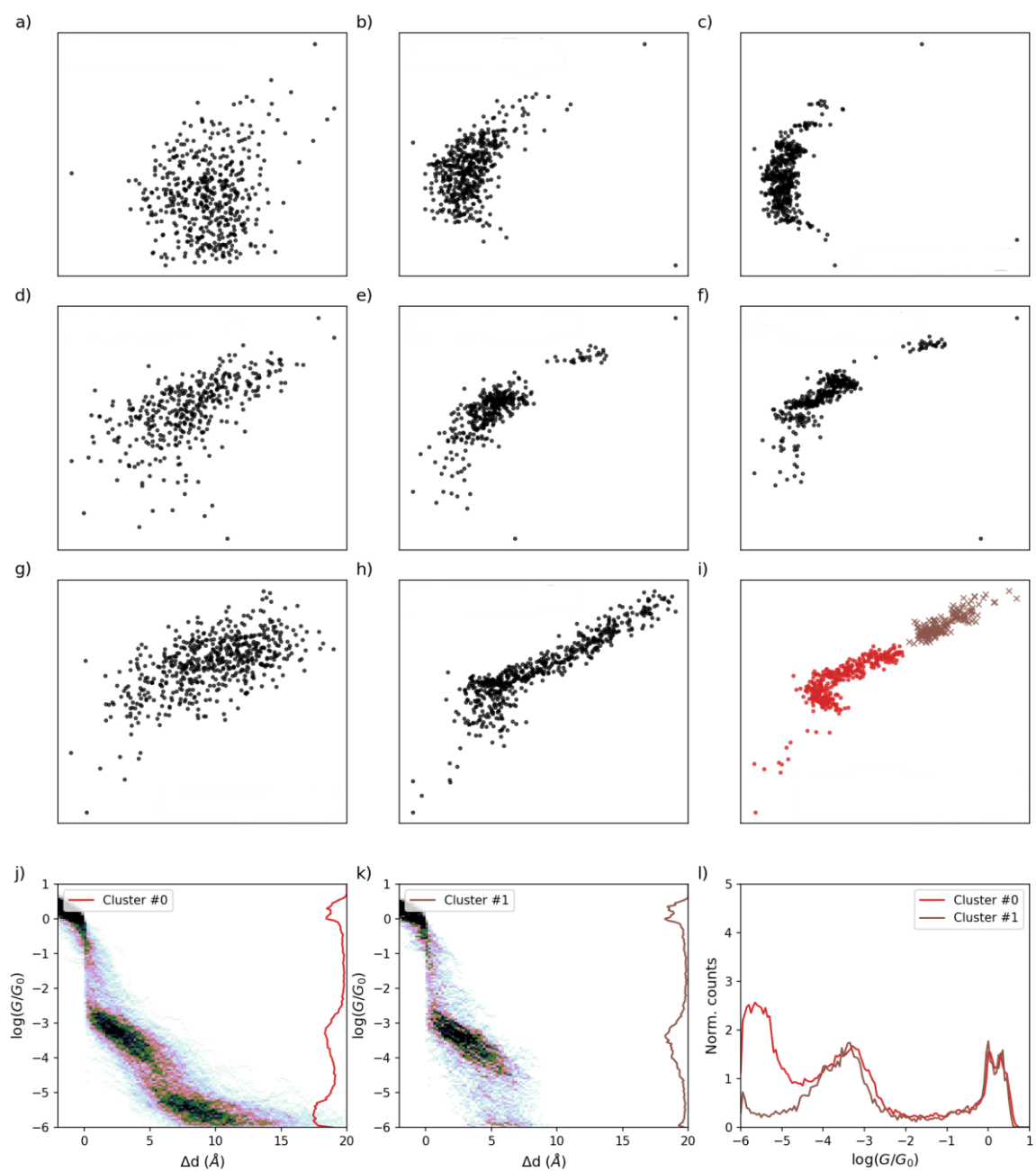

**Figure 4**. Effect of training time on the data classification. a-i) Scatter plots for the points learnt by the AE after 100 (left column), 300 (middle) and 1000 (right) epochs of training for all three molecules: BdNC (top row); MBdNC (centre) and tBuBdNC (bottom). In i) the two clusters determined by the spectral clustering algorithm are shown in colour (blue: cluster 0, orange: cluster 1). For tBuBdNC: j) Combined 2D/1D conductance histogram for the traces from cluster 0. k) Combined 2D/1D conductance histogram for the traces from cluster 1. l) Comparison of 1D conductance histograms for clusters 0 and 1.

Hence, in the final step of the analysis, we tried to identify for the MBdNC data as an example (cf. fig. 3), whether specific global features in the data traces have an effect on the classification. For this purpose, we modified the input data for cluster 1 mathematically and compared the separation between clusters 0 and 1 before and after the modification, cf. section 5 in the SI for further details. Specifically, this was done by A) removing the low-conductance region $\log(G/G_0) < -5.5$; B) adjusting the slope in the decay region ($\log(G/G_0) < -3$) to nominally the same value as in cluster 0; and C) by doubling the height of the conductance plateau in the region $-3 < \log(G/G_0) < -2$ (to the value observed for cluster 0). The high-conductance region is characterised by the formation of the Au/Au contact and is similar for clusters 0 and 1. It can therefore not be an important factor in the classification and was not considered further in the present analysis. As a quantitative measure for the cluster separation, we use the Fisher criterion, as shown in eq. (1).[34]

$$F = \frac{w^T S_B w}{w^T S_W w} \tag{1}$$

where $S_B$ and $S_W$ are the between-class covariance matrix and the within-class covariance matrix, respectively:

$$S_B = (m_0 - m_1)(m_0 - m_1)^T$$

$$S_W = \sum_{\text{cluster } 0}(x_i - m_0)(x_i - m_0)^T + \sum_{\text{cluster } 1}(x_j - m_1)(x_j - m_1)^T$$

and $w = S_W^{-1}(m_0 - m_1)$. $m_0$ and $m_1$ are the respective cluster means.

The results are shown in figure 5, namely the unmodified MBdNC dataset in panel a) (F = 0.066) and the datasets with modifications A), B) and C) in panels b) to d), respectively. Starting with case A, the removal of the low-conductance region ($\log(G/G_0) < -5.5$) in the traces belonging to cluster 1 leads an increase of the variance in this cluster as well as a minor shift away from cluster 0 (F = 0.099). Comparing this observation with the results shown in fig. 3 d) and e), this is at first glance surprising, given that the 2D histogram for cluster 1 shows a feature with high point density in this conductance region, while cluster 0 does not. It needs to be borne in mind, however, that the emergence of this feature is related to the relatively low variability of all parts of the traces in cluster 1, in comparison to cluster 0. Every individual trace still possesses a low-conductance region. Its removal reduces the similarity to the original, unmodified traces, while other global features are retained. This is why the association with cluster 1 becomes worse (cluster variance increases), but overall does not become more similar to traces in cluster 0.

A different picture emerges in case B), when the conductance traces are modified in such a way that the decay region ($\log(G/G_0) < -3$) is less steep and hence more similar to traces in cluster 0. Now all points in cluster 1 move towards cluster 0, fig. 5c, with substantial overlap between the two point groups. F decreases substantially from 0.066 to 0.003. Notably, the high-conductance region including the plateau feature at $-2 < \log(G/G_0) < -3$ remains unchanged during this operation.

Finally, we increased the average plateau height for traces in cluster 1 by a factor of 2 (case C). In this case, the decay and low-conductance regions remained unchanged, as in the original dataset. As before, the data points in cluster 1 on average move towards cluster 0, albeit to a lesser extent than in case B). The F-value decreases to 0.041.

Taken together, this assessment would suggest that the most important factor for the classification into the two groupings is indeed the slope of the decay region, followed by the plateau height. The low-conductance region appears to have the least influence on the overall classification result out of these three factors.

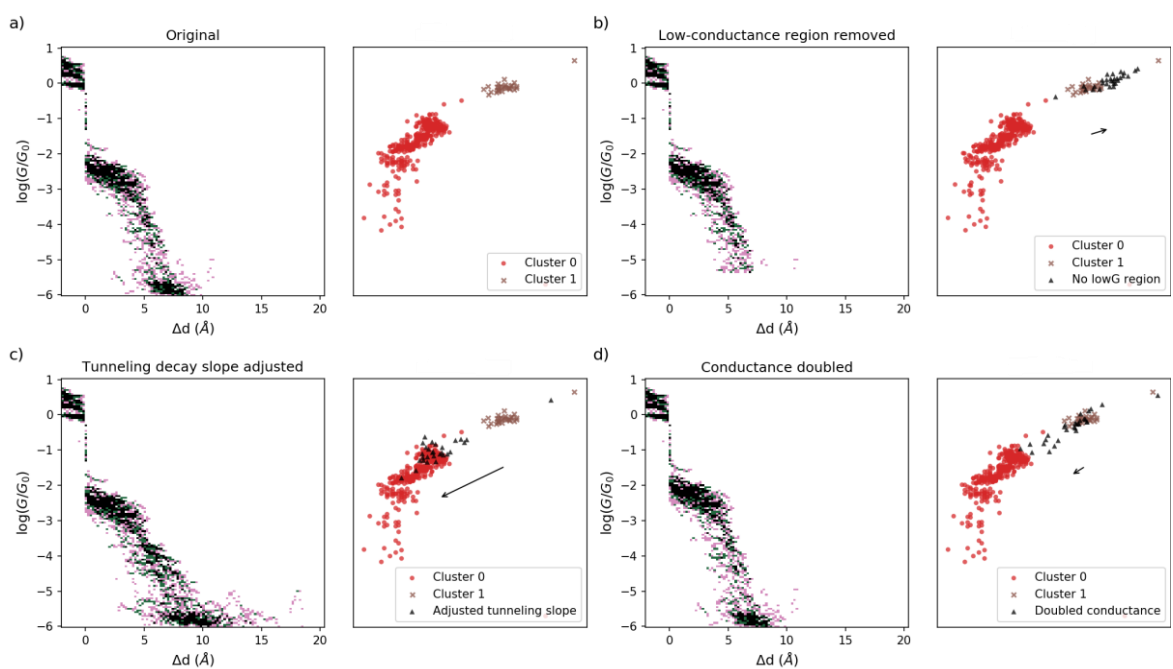

**Figure 5**: Effect of input data characteristics on classification results, MBdNC dataset. a) 2D histogram of original dataset from cluster 1. Clockwise, left side: 2D histograms after modifying three global aspects of the input data, namely the removal of the low-conductance region ($\log(G/G_0) < -5.5$, panel b)), after adjusting the slope in the decay region ($-5.5 < \log(G/G_0) < -3$, panel c)) and after doubling the height of the conductance plateau ($-3 < \log(G/G_0) < -2$, panel d)). On the right: Corresponding effect on the clustering result. Cluster 0 (red dots) remains unchanged in all cases; cluster 1 unmodified (brown crosses) and after modification (black triangles).

Finally, we return to the question of how to interpret the emergence of the two sub-populations for MBdNC and tBuBdNC, but not for BdNC, as opposed to a continuous distribution of molecular characteristics in all three cases. For MBdNC and tBuBdNC, we compared individual traces from the border region between the two clusters and found distinct differences in the steepness of the decay region, i.e. not a continuous variation. This supports the notion that the different clusters indeed represent distinct physical behaviour. To this end, our conclusions must remain somewhat speculative in the absence of further theoretical and/or experimental characterisation. However, it is well-known that longer and less steep decays in G(d) data can be associated with either structural reorganisation of the electrodes or lateral interactions between multiple molecules or sliding in the junction during the breaking process. Accordingly, steeper, better defined decay curves with a somewhat lower molecular conductance plateau could represent junctions formed by individual molecules. This could well be the case here: BdNC lacks bulky side groups and should therefore be more prone to lateral intermolecular interactions, making the formation of multi-molecule junctions with a wider and continuous distribution of junction properties more likely. On the other hand, side groups in both MBdNC and tBuBdNC could render this scenario less likely and the formation of well-defined single-molecule junctions could constitute a distinct sub-population in the data. Hence, in this specific case further work is needed for a more detailed, physical interpretation of the results. However, the ability of the TL/AE method to provide this level of detail is clearly of broader significance.

**Conclusions**

Here, we demonstrate an approach to unsupervised data classification, based on a TL methodology combined with dimensionality reduction. We exploit the capabilities and, importantly, prior training of openly available image recognition networks, and in particular AlexNet, to extract salient features in different types of single-molecule data (both simulated and experimental). We subsequently used AEs and other dimensionality reduction techniques to produce low-dimensional representations, as a basis for clustering and further analysis and interpretation. The combined AlexNet/AE approach performed well, compared to other dimensionality reduction techniques, such as PCA and AEs directly applied to our simulated and experimental data. Notably, the feature extractor did not rely on our own data for training, but rather used a large body of seemingly unrelated image data. This shows that training data for feature recognition can effectively be 'borrowed' from other applications, relaxing the need for large amounts of application-specific data. In the case of the MCBJ data shown above, we provide evidence for which global features formed the basis for the differentiation of data traces into different sub-populations. More generally, this depends on the ability of the feature extractor to recognise salient features in the data, against a background of non-distinctive features and noise, and of the dimensionality reduction component to subsequently preserve those features in the analysis.

**Methodology**

As mentioned in the main text, AlexNet can be separated into two main parts, the feature extractor and the classifier. The feature extractor, takes a 224x224 RGB image as an input and via the sequence of convolutional, relu and maxpool layers extracts `features' and generates 9216-dimensional feature vector for every image, which is then used for classification.

To apply the feature extractor directly to the molecular data, the latter must be presented as RGB images, or, in other words, as 3x224x224 numeric arrays. To do so, we generated tunnelling traces with 112x112 = 12544 points in each which were represented as 12544-dimensional vector. All values were then scaled into the range of [0,255], and line-by-line reshaped into 112x112 matrices. In the last step, matrices were upscaled to 224x224 size and repeated three times for R, G and B channels. The final representation of every simulated trace is 3x224x224 array of real values from the range between 0 and 255. All steps of the approach are shown in Figure 1 (a-c).

For dimensionality reduction of the extracted feature vectors, we designed an AE containing 7 hidden layers with 1024, 512, 128, 2, 128, 512, 1024 neurons, respectively (Fig.1 d). As an input and output we used layers with 9216 neurons, matching the output of the feature extractor of AlexNet. The weights of the code layer with two neurons were used for the two-dimensional representation of the data. For other networks, namely, VGG-16 and ResNet-18 we used the same architecture of the AE, except the number of neurons in the input and output layers was 25088, matching the dimensionality of the feature extractors of mentioned networks. All networks were implemented in Python programming language using the PyTorch deep learning framework. The weights of all neurons of all trained networks were taken from torchvision package.

The low-dimensional data obtained after training of the AE were used for visualisation as well as for clustering (Fig.1e). To compare the efficiency of proposed approach with classical machine learning and dimensionality reduction techniques, we used principal component analysis (PCA), t-stochastic neighbour embedding, multi-parameter vector classification (MPVC), and AE built on raw data. The preparation, processing and modification of the simulated nucleotide data and the experimental MCBJ data are described in detail in the SI.

**Acknowledgements**

A.V. and T.A. would like to thank the Artificial Intelligence and Augmented Intelligence for Automated Investigations for Scientific Discovery Network+ (AI3SD, EPSRC grant number: EP/S000356/1) and the Leverhulme Trust (RPG-2014-225) for funding, and Prof. Michel Calame for comments on an earlier draft of the manuscript.**References**

[1] Alsteens, D., Gaub, H.E., Newton, R., Pfreundschuh, M., Gerber, C., Muller, D.J., Atomic force microscopy-based characterization and design of biointerfaces, Nat. Rev. Mater. **2**, 17008 (2017).

[2] Pujals, S., Feiner-Gracia, N., Delcanale, P., Voets, I., Albertazzi, L., Super-resolution microscopy as a powerful tool to study complex synthetic materials, Nat. Rev. Chem. **3**, 68-84 (2019).

[3] Sahl, S.J., Hell, S.W.; Jakobs, S., Fluorescence nanoscopy in cell biology, Nat. Rev. Mol. Cell Biol. **18**, 685-701 (2017).

[4] Nichols, R.J., Higgins, S.J., Single Molecule Nanoelectrochemistry in Electrical Junctions, Acc. Chem. Res. **49**, 2640-2648 (2016).

[5] Albrecht, T., Electrochemical tunnelling sensors and their potential applications, Nat. Commun **3**, 829 (2012).

[6] Inkpen, M.S., Mario Lemmer, M., Fitzpatrick, N., Costa-Milan, D., Nichols, R.J., Long, N.J., Albrecht, T., New insights into single-molecule junctions using a robust, unsupervised approach to data collection and analysis, J. Amer. Chem. Soc. **137**, 9971-9981 (2015).

[7] Frei, M., Aradhya, S.V., Hybertsen, M.S., et al., Linker Dependent Bond Rupture Force Measurements in Single-Molecule Junctions, J. Amer. Chem. Soc. **134**, 4003-4006 (2012).

[8] Reuter, M. G., Hersam, M. C., Seideman, T., Ratner, M. A., Signatures of cooperative effects and transport mechanisms in conductance histograms. Nano Lett. **12**, 2243-2248 (2012).

[9] Lemmer, M., Inkpen, M. S., Kornysheva, K., Long, N. J., Albrecht, T., Unsupervised vector-based classification of single-molecule charge transport data. Nat. Commun. **7**, 12922 (2016).

[10] Hamill, J.M., Zhao, X.T., Mészáros, G., Bryce, M.R., Arenz, M.,Fast Data Sorting with Modified Principal Component Analysis to Distinguish Unique Single Molecular Break Junction Trajectories, Phys. Rev. Lett. **120**, 016601 (2018).

[11] Cabosart, D., El Abbassi, M., Stefani, D., Frisenda, R., Calame, M., van der Zant, H.S.J., Perrin, M.L., A reference-free clustering method for the analysis of molecular break-junction measurements, Appl. Phys. Lett. **114**, 143102 (2019).14

# Unsupervised Classification of Single-Molecule Data with Autoencoders and Transfer Learning

Anton Vladyka[1], Tim Albrecht[1]

[1] School of Chemistry, University of Birmingham, Edgbaston Campus, Birmingham B15 2TT, United Kingdom

E-mail: t.albrecht@bham.ac.uk

# Supporting Information

**1) Extracting features using trained artificial neural networks**

Implementation of the AlexNet neural network as well as the weights for the network trained on Imagenet dataset is available via torchvision package, or from

https://pytorch.org/docs/stable/_modules/torchvision/models/alexnet.html#alexnet

For convenient access of the pretrained AlexNet, the weights of AlexNet can be downloaded from https://download.pytorch.org/models/alexnet-owt-4df8aa71.pth as pytorch state dictionary, and saved as a separate file.

As an input, AlexNet takes 224x224 RGB images, or in terms of pytorch, real-valued tensors with a shape of 1x224x224x3.

To load the model with pretrained weights:

```
def alexnet(pretrained=False, **kwargs):
    r"""AlexNet model architecture from the
    `"One weird trick..." <https://arxiv.org/abs/1404.5997>`_ paper.
    Args:        pretrained (bool): If True, returns a model pre-trained on ImageNet
     """
    model = AlexNet(**kwargs)
    if pretrained:
        model.load_state_dict(torch.load('<path_to_pth_file>/alexnet-owt-4df8aa71.pth'))
    return model
```

To get access to the features extracted by the AlexNet feature extractor, an extra method had to be added to the AlexNet class:

```
class AlexNet(nn.Module):
    …
    …
    def getfeatures(self, x):
        x = self.features(x)
        x = x.view(x.size(0), 256 * 6 * 6)
        return x
```

This method returns a pytorch tensor with a shape of 9216x1.

Python implementations of other studied networks, ResNet-18 and VGG-16, are available from https://pytorch.org/docs/stable/_modules/torchvision/models/resnet.html#resnet18 and https://pytorch.org/docs/stable/_modules/torchvision/models/vgg.html#vgg16, respectively, and both networks were used in the same manner. Both networks accept 1x224x224x3 arrays as an input, but output 25088-dimensional feature vectors.

**2) Data preparation**

2.1 Nucleotides tunnelling traces simulation

Simulated tunnelling traces I(t) for different nucleotides were generated using the same parameters as in [1]:

| Class | E | $m_d/\sigma_d$ | $(m_m/\sigma_m)*0.01$ | $(m_n/\sigma_n)*0.001$ | $p_s$ |
|---|---|---|---|---|---|
| Noise | - | - | - | 2.0/1.0 | 0 |
| A | 40 | 150/50 | 2.5/0.5 | 2.0/1.0 | 0.3 |
| C | 35 | 25/15 | 4.5/0.5 | 2.0/1.0 | 0 |
| G | 25 | 40/15 | 5.0/2.0 | 2.0/1.0 | 0 |
| T | 65 | 50/15 | 4.0/4.0 | 2.0/1.0 | 0.05 |
| MetC | 55 | 5/15 | 2.0/0.5 | 2.0/1.0 | 0 |

where E is a number of events per 10000 data points, $m_d$ and $\sigma_d$ are mean and standard deviation of the duration of all events in data points, respectively, $m_m$ and $\sigma_m$ are mean and standard deviation of the event magnitude, respectively, $m_n$ and $\sigma_n$ are mean and standard deviation of the noise fluctuations, $p_s$ is the probability to observe telegraphic switching between event magnitude level and baseline level within the event. Event duration, event magnitude and noise fluctuation are generated from the normal distribution with the mean and standard deviation, as specified above:

Event duration $\sim N(m_d, \sigma_d)$

Event magnitude $\sim N(m_m, \sigma_m)$

Baseline = noise $\sim N(m_n, \sigma_n)$

2.2 Converting the nucleotide data into images

We generated 2000 traces/class for all 6 classes. For convenience, all traces were generated with 112*112 = 12544 points keeping the same frequency of events as in above table. After generation, each array of the shape 12544x1 was reshaped into a square 112x112 array line by line, upscaled to 224x224, and then extended into the third dimension to get pseudo-grayscale image.

2.3 Converting the RBdNC data into images (RBdNC = BdNC, MBdNC or tBuBdNC)

Raw data for all the molecules were taken as a sequence of 224 points (d, G), where d is a displacement in Ångstrom, from -2 to 20.3, and G is the conductance in units of $\log(G/G_0)$, in a range between -6 (G = $10^{-6}G_0$, noise level) and 1 (G=$10G_0$).

To convert conductance data into a graphical representation, each trace was binned in two dimensions with 112 bins in each direction. After binning, each trace was given as a 112x112 sparse matrix, then upscaled to 224x224 and replicated in the third dimension to get the shape of 224x224x3. In this way, the data were exactly in the right format to be used as input, e.g. in AlexNet.

### 3) Autoencoders for dimensionality reduction

All AEs used for dimensionality reduction, have the same symmetric 9-layer structure (figure S1), where input and output layers have the same number of neurons as the number of features extracted by the feature extractor of studied trained neural network (9216 for AlexNet, 25088 for ResNet18 and VGG-16, 12544 neurons for raw data), and $5^{th}$ (code) layer has 2 neurons. The activations of the neurons in code layer are interpreted as (x,y) coordinates for two-dimensional representation of data after dimensionality reduction.

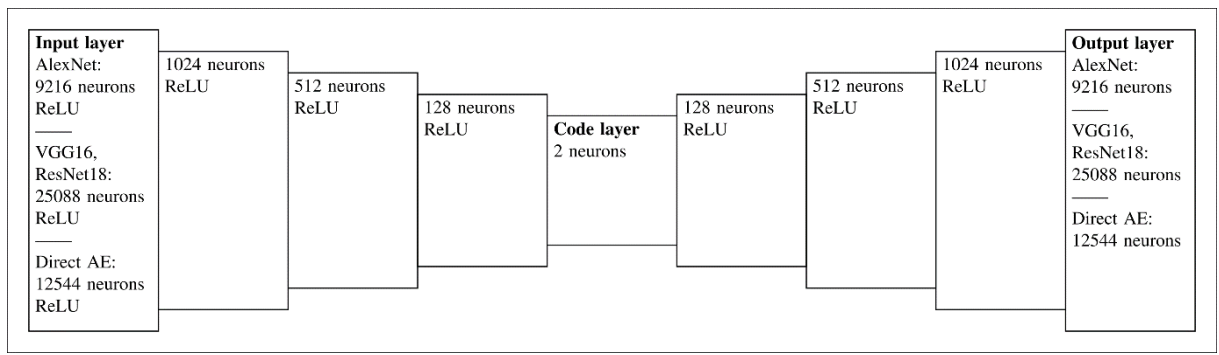

**Figure S1**. Schematics of the autoencoder used in this research. Depending on the netwoorks used for feature extraction, input and output layers have 9216 neurons for AlexNet and 25088 neurons for VGG-16 and ResNet-18. For the autoencoder-based dimensionality reduction without use of trained neural network, 112x112 = 12544 neurons in input and output layers were used for the nucleotide study, and 224 neurons in input and output layers for the RBdNC study.

### 4) Classical approaches for dimensionality reduction

We also compared the Transfer Learning methodology with 'classical' approaches for dimensionality reduction, such as principal component analysis (PCA), t-distributed stochastic neighbour embedding (tSNE, [4]), autoencoder (AE) on raw data ('direct AE'), multi-parametric vector classifier (MPVC, [3]).

PCA is a classic approach to dimensionality reduction and an implementation of PCA in python is available via scikit-learn toolkit.[2] Data were analysed 'as is', without prior modification or normalisation. The first two principal components were used for representation.

tSNE is a stochastic method of dimensionality reduction, which is trying to equalize the conditional probabilities for the points to be similar to each other in the original high-dimensional space and in desired low-dimensional one.[4] Implementation of tSNE in Python is available via scikit-learn toolkit. The results of tSNE depend on its `perplexity' parameter, which is an estimate of the number of neighbours from the same class each point has. We run the algorithm with different perplexities from 1 to 100.

MPVC is a machine learning algorithm for dimensionality reduction where the features to characterize every point in multidimensional space are defined manually. MPVC of simulated nucleotide traces was implemented following [3]. As a reference **R**, the median of all simulated traces was used. For every simulated conductance trace $\mathbf{X_m}$, the length of the difference vector $|\mathbf{Y_m}| = |\mathbf{X_m}-\mathbf{R}|$ was calculated together with an angle $\theta_m$ between $\mathbf{Y_m}$ and $-\mathbf{R}$. Calculated parameters $|\mathbf{Y_m}|$ and $\theta_m$ were used for two-dimensional representation of each simulated trace. For RBdNC data, an average trace of all 1425 traces was used as a reference **R**.

We analysed the results for the overall dataset as well as for each individual molecule (Fig. S2). The black points represent those identified as cluster 1 in the TL approach (for MBdNC), cf. main text. None of the above algorithms was able to separate the overall dataset into sub-populations according to the three molecules or identified sub-populations within a given molecular dataset. Nevertheless, traces from cluster 1 appear in roughly the same region of the cluster plot, as indicated by the black data points. This suggests that the corresponding traces are being recognised as similar, even if they do not form separate clusters.

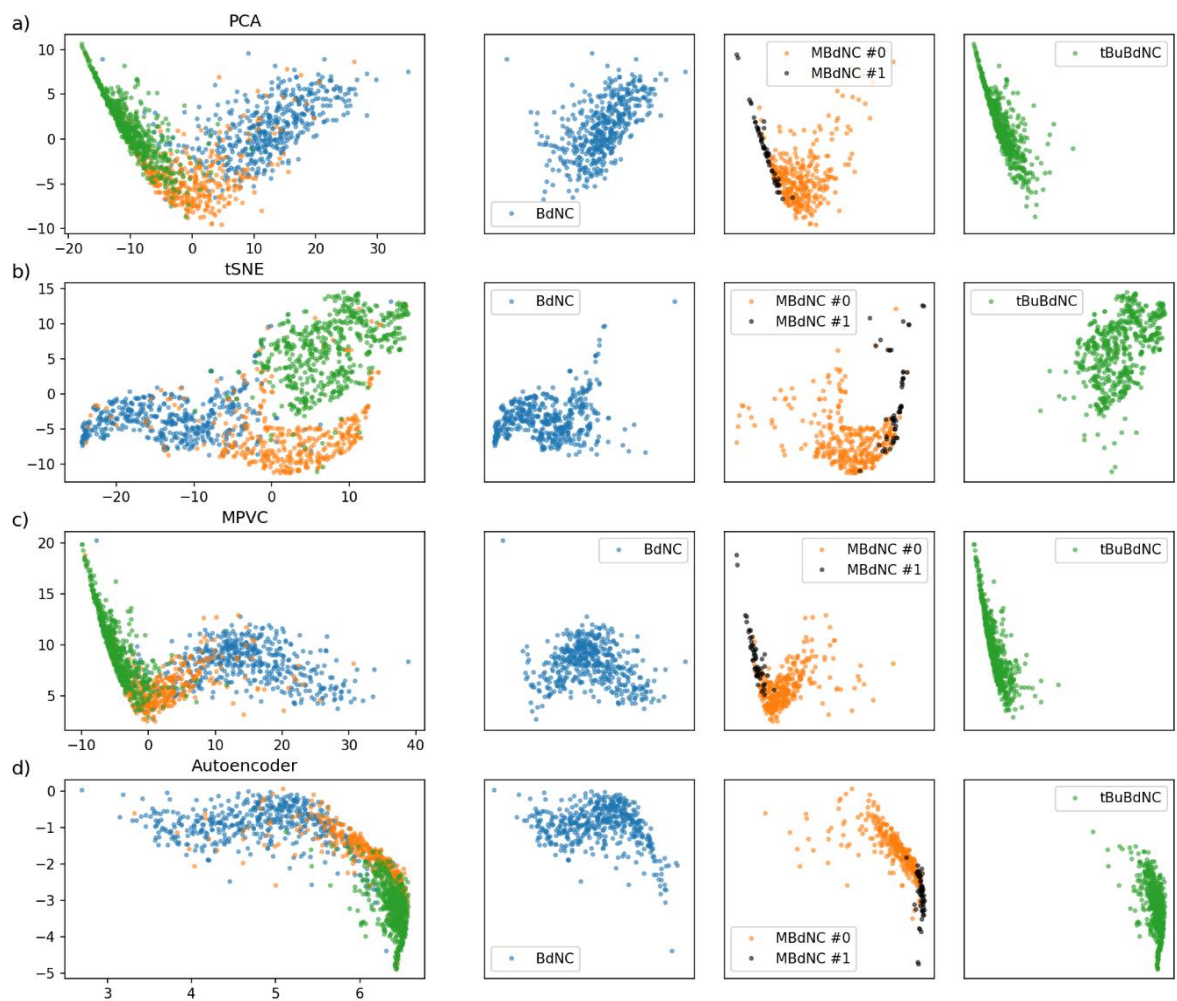

**Figure S2**. Classical dimensionality reduction approaches applied to raw RBdNC data: 1425 G(d) traces G(d) with 224 points in each. a)-d) Results for PCA, tSNE (perplexity=40), MPVC, Autoencoder. For the MBdNC dataset, the subpopulation of cluster #1 determined by TL approach is shown in black.

**5) Identification of the features which define separation of the clusters for MBdNC**

To identify which features in the traces are likely to be the basis for classification, we defined three regions-of-interests (ROI) in the conductance-displacement histograms which are different for clusters 0 and 1. Then, each trace in the cluster 1 was artificially transformed in a certain way, and then passed through the feature extracted of the AlexNet and AE (see Figure 5 in the main text). The ROIs are the corresponding transformations of the traces are the following:

a) Low-conductance region: removed, i.e. each point of the trace below -5.5 placed out of range (below -6.5).
b) Slope of tunneling decay: adjusted to match the slope for the traces from cluster 0.
c) Conductance plateau: plateau shifted up mimicking increase of the conductance by a factor of 2, or increased by +0.3 at the logarithmic scale. The position of low-conductance region remains the same.

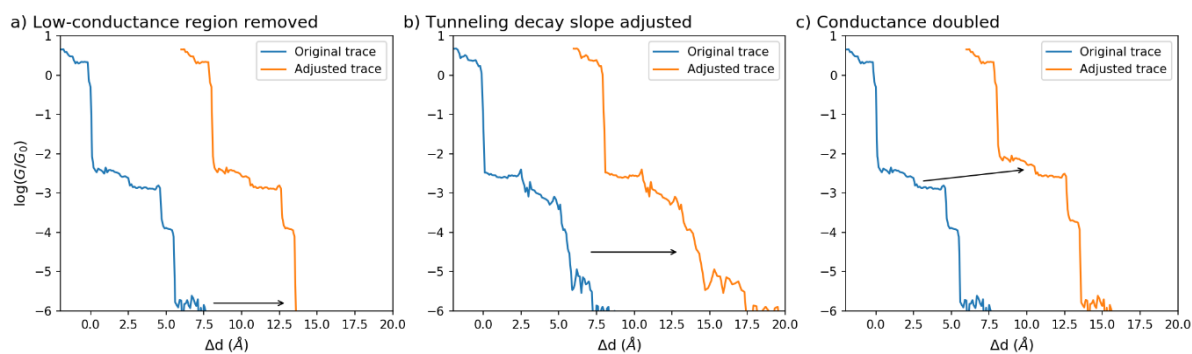

**Figure S3**. Transformations of the traces used to identify the role of the individual features in classification. a) Removing the low-conductance feature. b) Adjusting the slope of the tunnelling region. c) Increasing the conductance of the plateau region.